\documentclass[aps,prd,twocolumn,showpacs,10pt,superscriptaddress,preprintnumbers,nofootinbib]{revtex4-1}
\usepackage{epsfig,amssymb,amsmath,psfrag,epstopdf,color}
\usepackage{graphicx}
\usepackage{collref}
\usepackage{multirow}

\allowdisplaybreaks
\def\Ord{{\cal O}}
\def\sss{\scriptscriptstyle}
\newcommand{\dd}{\mathrm{d}}
\newcommand{\nbrk}{\nonumber\\}

\begin{document}

\preprint{NSF-KITP-16-077}
\preprint{MIT-CTP-4807}

\title{NNLO QCD Corrections to $t$-channel Single Top-Quark
    Production and Decay}
\author{Edmond L. Berger}
\email{berger@anl.gov}
\affiliation{High Energy Physics Division,
Argonne National Laboratory, Argonne, Illinois 60439, USA}
\author{Jun Gao}
\email{jgao@anl.gov}
\affiliation{High Energy Physics Division,
Argonne National Laboratory, Argonne, Illinois 60439, USA}
\affiliation{Kavli Institute for Theoretical Physics,
University of California, Santa Barbara, CA 93106, USA}
\author{C.-P. Yuan}
\email{yuan@pa.msu.edu}
\affiliation{Department of Physics and Astronomy,
Michigan State University, East Lansing, Michigan 48824, USA}
\author{Hua~Xing~Zhu}
\email{zhuhx@mit.edu}
\affiliation{Center for Theoretical Physics, Massachusetts
  Institute of Technology, Cambridge, MA 02139, USA}

\begin{abstract}
\noindent
We present a fully differential next-to-next-to-leading order calculation of
$t$-channel single top-quark production and decay at the LHC under narrow-width
approximation and neglecting cross-talk between incoming protons. 
We focus on the fiducial cross sections at 13 TeV, finding that the next-to-next-to-leading
order QCD corrections can reach the level of $-6\%$.
The scale variations are reduced to the level of a percent.
Our results can be used to improve experimental acceptance estimates and 
the measurements of the single top-quark
production cross section and the top-quark electroweak couplings. 
\end{abstract}

\pacs{}
\maketitle

\section{Introduction}
The top quark can be produced singly at a hadron collider through the
electroweak~(EW) $Wtb$ vertex. There are three production channels: 
the $t$-channel and $s$-channel processes through exchange of a $W$ boson, 
and associated production of $tW$.  All three channels are sensitive to the structure of 
the $Wtb$
vertex and to the CKM matrix element $V_{tb}$, an important motivation for their study.
Moreover, single top production provides an important
window to physics beyond the standard
model~(SM)~\cite{hep-ph/0007298}, e.g., a modified $Wtb$ vertex, new
heavy quarks, new gauge bosons, flavor-changing neutral current, and so forth.  
Single top-quark production was first
established at the Fermilab Tevatron~\cite{0903.0885,0903.0850}, and later at
the Large Hadron Collider~(LHC)~\cite{PHLTA.B717.330,PHLTA.B716.142,Chatrchyan:2012ep,PRLTA.110.022003}.
Single top-quark studies are expected to enter an era of high precision 
during the upcoming run of the LHC at higher energy and larger luminosity.

The $t$-channel production has the largest rate among the three 
at the LHC, about $210$ pb at $\sqrt{s}=13$ TeV.  
Significant efforts to improve the theoretical description of this
process include 
next-to-leading order (NLO) QCD corrections
in both 4- and 5-flavor schemes 
in Refs.~\cite{NUPHA.B435.23,hep-ph/9603265,hep-ph/9705398,hep-ph/0207055,Sullivan:2004ie,hep-ph/0504230,hep-ph/0408158,1007.0893,1012.5132}.  Soft
gluon resummation has been considered in
Refs.~\cite{1010.4509,1103.2792,1210.7698,1510.06361}.  Matching NLO
calculations to parton showers is done in
Refs.~\cite{hep-ph/0512250,0907.4076,1207.5391}.  Recently, next-to-next-to-leading order (NNLO) QCD
corrections with a {\em stable} top quark were calculated in
Ref.~\cite{1404.7116}, neglecting certain subleading contributions in color.

In this manuscript we present the first fully differential NNLO calculation of $t$-channel single top quark 
{\em production and decay} at the LHC in the 5-flavor scheme in QCD.  We follow 
Ref.~\cite{1404.7116} in neglecting cross-talk between the hadronic systems of the two incoming 
protons.  To the best of our knowledge, our calculation is the first to combine QCD corrections at NNLO 
for production and decay, meaning that a 
realistic simulation at NNLO is now available for leptonic top-quark decay in 
$t$-channel single top-quark production.  The differential nature of our calculation 
allows us to impose phase space selections on final state objects, as done in the experiments 
(fiducial cuts).  Using the fiducial cuts of the ATLAS and CMS 
analyses~\cite{ATLAS:2014dja,CMS:2015jca}, we find 
large NLO corrections, thus necessitating the higher order calculation performed here.  The 
uncertainties associated with QCD hard-scale variation are reduced to the level of $\sim 1\%$.   
We compute the ratio of the top anti-quark to top quark production distributions with fiducial cuts 
applied, showing that sensitivity of this charge ratio to current parton distribution functions is much 
larger than to the perturbative corrections.  

In the remaining paragraphs we outline the methods used in the calculation, and we
present our numerical results on inclusive cross sections and fiducial
cross sections and various differential distributions.

\section{Method}\label{sec:med}
\begin{figure}[!h]
  \begin{center}
  \includegraphics[width=0.35\textwidth]{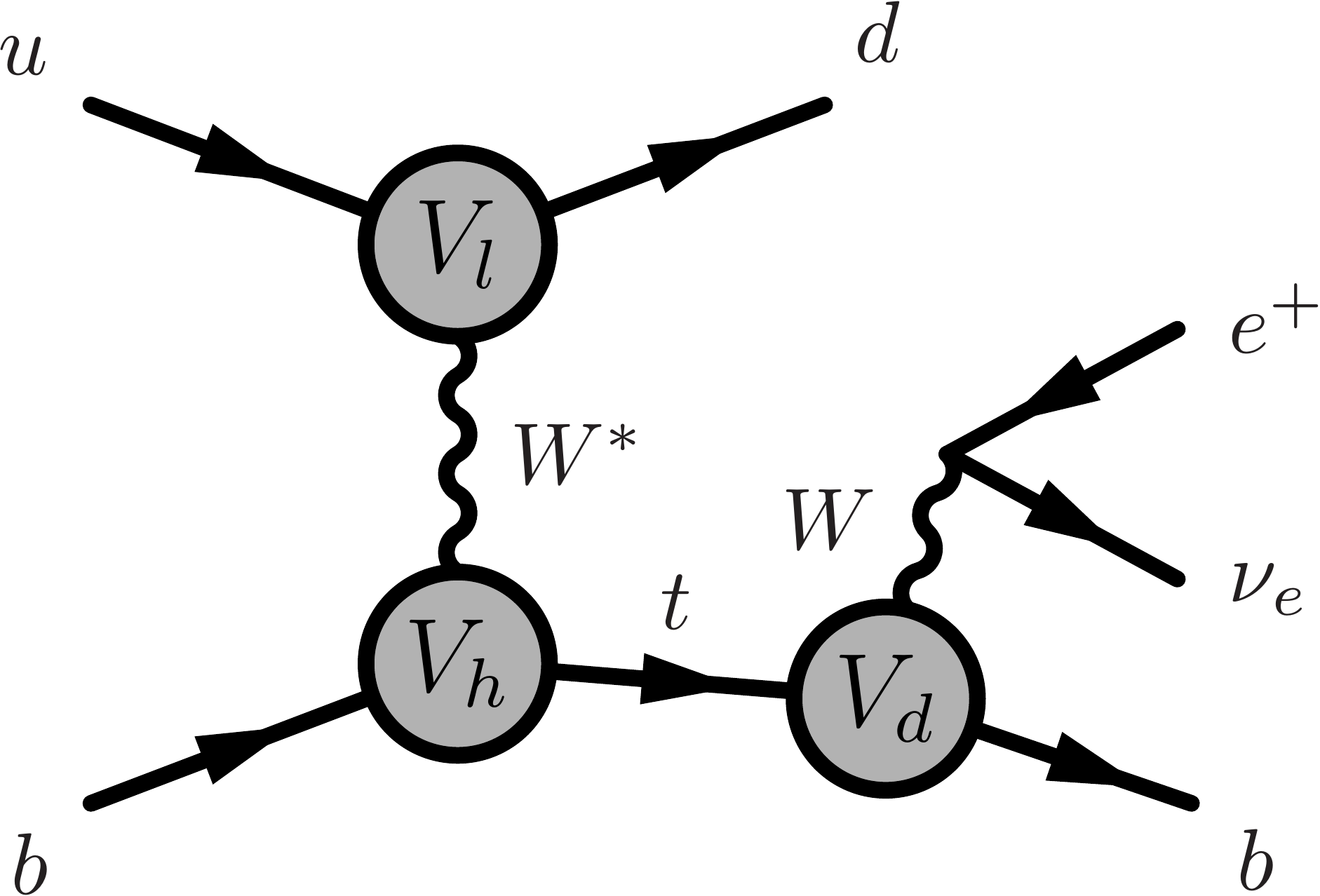}
  \end{center}
  \vspace{-2ex}
  \caption{\label{fig:singletop}
  Sketch of t-channel single-top quark production and decay; $u b \rightarrow d t$ with $t \rightarrow e^+ \nu_e b$.  
  $V_l$ represents QCD corrections to the light quark line, which could include interference of the tree-diagram and the two-loop diagram, square of the one-loop diagram (double-virtual), interference of the one-loop diagram with one additional gluon and the tree-level diagram with one additional gluon (real-virtual), and 
the square of tree-level diagram with a pair of additional partons (double-real).  $V_h$ and $V_d$ represent the same type of corrections to the heavy quark line and the decay part, respectively. There is no cross talk between the light quark line, heavy quark line, and the decay part in our calculation.}
\end{figure}

We work in the Narrow-Width Approximation~(NWA), under which the QCD corrections
to the top-quark production and decay are factorizable.  
As confirmed by explicit numerical studies of off-shell and
non-factorizable corrections~\cite{1102.5267,1305.7088}, this 
approximation is justified because the top-quark decay width
is much smaller than its mass.  
We approximate the full QCD corrections by the
\emph{vertex} corrections; in the inclusive case, this is known
as the structure function approach~\cite{hep-ph/9206246}.  In this
approach, we separate the QCD corrections into three factorizable
contributions, as sketched in Fig.~\ref{fig:singletop}.  
There is a decay
part $V_d$; the light-quark line of the production part $V_l$; and the heavy-quark
line of the production part $V_h$. The use of the (fully differential) structure function
approximation 
builds on the observation that
QCD interference between the light-quark line and the heavy-quark line
vanishes at NLO and is suppressed at least by a factor of $1/N_c^2$ at
NNLO.  The use of such an approximation is also crucial for making this
calculation feasible, because interference contributions between 
the light and heavy-quark lines are not yet available~\cite{1409.3654} 
for the full two-loop virtual diagrams.  The structure function
approximation at NNLO is also used in an earlier calculation of
$t$-channel on-shell single top-quark production~\cite{1404.7116}, and in Higgs boson production through weak boson fusion~\cite{1003.4451,1506.02660}. 

The NNLO QCD corrections to the heavy-quark line are straightforward if we use
phase-space slicing with the $N$-jettiness variable~\cite{Stewart:2010tn,Boughezal:2015dva,
Gaunt:2015pea}.
A similar calculation was performed for charm quark
production in neutrino deep inelastic scattering (DIS) in Ref.~\cite{Berger:2016inr}.    
For the corrections to the light-quark line, we adopted the
method of ``projection-to-Born'' in Ref.~\cite{1506.02660}.  The key
ingredients in this approach are the inclusive NNLO DIS coefficient
functions~\cite{vanNeerven:1991nn,Zijlstra:1992qd,Zijlstra:1992kj}, for
which a conveniently parametrized version is
available~\cite{vanNeerven:1999ca,vanNeerven:2000uj}.   For the 
real-virtual corrections, we extracted the one-loop helicity amplitudes
from DIS 2 jet production in Ref.~\cite{0904.2665}, and we cross checked 
with Gosam~\cite{1404.7096}.  These ingredients were combined according to
Ref.~\cite{1506.02660}, by constructing appropriate counter-events with
opposite weights for every event in the Monte Carlo (MC) integration of double-real and real-virtual
contributions, which render the phase space integrals finite for infrared (IR) safe observables.
For the decay part of the calculation, we adopted the results in 
Ref.~\cite{1210.2808}.   We also take into account the product of two NLO corrections 
from different combinations of the light-quark line, the heavy-quark line, and the decay part.

Finally, we combine corrections to the production part and
decay part consistently in the NWA, as in 
Refs.~\cite{hep-ph/0403035,0907.3090,1407.2532}.  Schematically, we write 
\begin{align}
  \label{eq:nwa}
  \sigma^{\rm LO} = &\,  \frac{1}{\Gamma^0_t} \dd \sigma^0 \dd
                      \Gamma^0_t
\nbrk
\delta\sigma^{\rm NLO} = & \,  \frac{1}{\Gamma^0_t}\Big[\dd \sigma^1 \dd
                     \Gamma^0_t+  \dd \sigma^0(\dd
                     \Gamma^1_t -
                     \frac{\Gamma^1_t}{\Gamma^0_t} \dd \Gamma^0_t)\Big]
\nbrk
\delta\sigma^{\rm NNLO} = & \,  \frac{1}{\Gamma^0_t}\Big[ \dd \sigma^2 \dd
                     \Gamma^0_t
+ \dd \sigma^1 (\dd
                     \Gamma^1_t-\frac{\Gamma^1_t}{\Gamma^0_t}\dd \Gamma^0_t)
\nbrk & \hspace{-0.2in}
+ \dd \sigma^0 (\dd
                     \Gamma^2_t
 - \frac{\Gamma^2_t}{\Gamma^0_t} \dd \Gamma^0_t  - 
\frac{\Gamma^1_t}{\Gamma^0_t} (\dd \Gamma^1_t-\frac{\Gamma^1_t}{\Gamma^0_t}\dd \Gamma^0_t))\Big],
\end{align}
where $\dd\sigma^i$ and $\dd\Gamma^i_t$ denote the $\Ord(\alpha_{\sss
  S}^i)$ corrections to the production and decay parts, respectively.  For 
the full NNLO corrections there  are contributions from
$\mathcal{O}(\alpha^2_{\sss S})$ production only,
from the product of $\mathcal{O}(\alpha_{\sss S})$ production and
$\mathcal{O}(\alpha_{\sss S})$ decay, and from $\mathcal{O}(\alpha^2_{\sss S})$ decay
only, as shown in Eq.~(\ref{eq:nwa}).
Inclusive production cross sections at each order can be obtained after integration over the decay phase space. 

\section{Numerical results} 
We use a top quark mass of 173.2 GeV and a $W$ boson mass of
80.385 GeV.  We set the $W$ boson decay branching ratio to 0.1086 for 
one lepton family.  We choose $|V_{tb}|=1$ and the CT14 NNLO parton
distribution functions (PDFs)~\cite{Dulat:2015mca}
with $\alpha_s(M_Z)=0.118$. The nominal central scale choice is $\mu_R=\mu_F=m_t$ with
scale variation calculated by varying the two together over the range $0.5 < \mu/\mu_o < 2$.
We list the LO, NLO and NNLO results for top quark and anti-quark production
in Table.~\ref{tab:total}. The NNLO QCD corrections reduce the cross sections
by $2\sim 3$~\% compared to a reduction of $4\sim 5$~\% at NLO. 
The full NNLO corrections consist of
pieces from the heavy-quark line, the light-quark line, and the products of them.  
There are cancellations
between these pieces as well as cancellations among different partonic channels.  
Perturbative convergence of the separate QCD series is well maintained, as we verified by 
checking the individual pieces.  Variations of the theoretical results associated with 
choices of the hard scales are reduced by a factor of 4 at NLO compared with LO, and 
by a further factor of 3 at NNLO with respect to NLO.
\begin{table}[h!]
\centering
\begin{tabular}{l|c|c|c} \hline
inclusive [pb]  & LO & NLO & NNLO  \\  [1ex] 
\hline \hline 
$t$ quark      & $143.7_{-10\%}^{+8.1\%}$ &  $138.0_{-1.7\%}^{+2.9\%}$ & $134.3_{-0.5\%}^{+1.0\%}$   \\  [1ex] 
\hline
$\bar t$ quark & $85.8_{-10\%}^{+8.3\%}$ & $81.8_{-1.6\%}^{+3.0\%}$ & $79.3_{-0.6\%}^{+1.0\%}$  \\  [1ex] 
\hline
\end{tabular}
\caption{Inclusive cross sections for top (anti-)quark production at 
13 TeV at various orders in QCD.  The scale uncertainties are calculated
by varying the hard scale from $\mu_F=\mu_R=m_t/2$ to $2\,m_t$, and are 
shown in percentages.}
\label{tab:total}
\end{table}

Fiducial cross sections for $t$-channel single top-quark production
have been measured at 7 and 8 TeV~\cite{ATLAS:2014dja,CMS:2015jca}.
We choose a fiducial region for 13 TeV that is similar to the one used
in the CMS analysis~\cite{CMS:2015jca} at 8 TeV.  We use the anti-$k_T$ jet 
algorithm~\cite{Cacciari:2008gp} with a distance parameter $D=0.5$.  Jets are defined 
to have transverse momentum $p_T>40$ GeV and pseudorapidity $|\eta|<5$.  We require exactly two jets in the final state,  
following the CMS and ATLAS analyses, meaning that events with additional jets are 
vetoed, and we require at least one of these to be a $b$-jet with $|\eta|<2.4$~\cite{foota,Banfi:2006hf}.
We demand the charged lepton to have a $p_T$ greater
than 30 GeV and rapidity $|\eta|<2.4$.  For the fiducial cross sections reported below we
include top-quark decay to only one family of leptons.  
 
Table~\ref{tab:fiducial} shows our predictions of the fiducial cross sections
at different perturbative orders, with scale variations shown in
percentages.
We vary the renormalization and factorization scales $\mu_R=\mu_F$ in the top-quark 
production stage, and the renormalization scale in the decay stage, 
independently by a factor of two around the nominal central scale choice.
The resulting scale variations are added in
quadrature to obtain the numbers shown in Table~\ref{tab:fiducial}. We also
show the QCD corrections from production and decay separately as defined in
Eq.~(\ref{eq:nwa}).  All results shown in Table~\ref{tab:fiducial} are for the
central scale choice $m_t$,  as for the inclusive cross sections.
The NNLO corrections from the product of $\mathcal{O}(\alpha_{\sss S})$
production and $\mathcal{O}(\alpha_{\sss S})$ decay can be derived by subtracting the 
above two contributions from the full NNLO corrections.  

Changes of the QCD corrections after all kinematic cuts are applied are evident if one  
compares Table~\ref{tab:fiducial} with Table~\ref{tab:total}.  Acceptance in the charged lepton, 
the $b$-jet, and the non-$b$ jet produce these changes, as well as the jet 
veto.  We call attention to the fact that the NLO QCD corrections in production have changed 
to $-19$\%.  The NLO corrections in decay further reduce the cross sections
by about 8\%.  At NNLO the correction in production is still dominant and can
reach $-6$\%.  The size of the NNLO correction in decay is smaller by about a factor of 2, and it almost
cancels with the correction from the product of one-loop production and one-loop decay.
Scale variations have been reduced to about $\pm 1\%$ at
NNLO. Scale variation bands at various orders do not overlap with each other in general.

\begin{table}[h!]
\centering
\begin{tabular}{l|l|c|c|c} \hline
\multicolumn{2}{c|}{fiducial [pb]}  & LO & NLO & NNLO  \\  [1ex] 
\hline \hline 
\multirow{3}{*}{$t$ quark} & total & $4.07_{-9.8\%}^{+7.6\%}$ &  $2.95_{-2.2\%}^{+4.1\%}$ & $2.70_{-0.7\%}^{+1.2\%}$   \\  [1ex] 
  & corr. in pro. &  & -0.79 & -0.24  \\ [1ex]
  & corr. in dec. &  & -0.33 & -0.13  \\ [1ex]
\hline
\multirow{3}{*}{$\bar t$ quark} & total & $2.45_{-10\%}^{+7.8\%}$ & $1.78_{-2.0\%}^{+3.9\%}$ & $1.62_{-0.8\%}^{+1.2\%}$  \\  [1ex] 
  & corr. in pro. &  & -0.46 & -0.15  \\ [1ex]
  & corr. in dec. &  & -0.21 & -0.08  \\ [1ex]
\hline
\end{tabular}
\caption{Fiducial cross sections for top (anti-)quark production with decay at 13 TeV 
at various orders in QCD with a central scale choice
of $m_t$ in both production and decay. The scale uncertainties correspond to
a quadratic sum of variations from scales in production and decay, and
are shown in percentages. Corrections from pure production and decay
are also shown.}
\label{tab:fiducial}
\end{table}

With fiducial cuts applied, the jet veto introduces another hard scattering
scale of $p_{T,veto}= 40$ GeV in addition to $m_t$.  Thus it may be appropriate
to choose a QCD scale of $(p_{T,veto}m_t)^{1/2}\sim m_t/2$, especially at lower 
perturbative orders where the gluon splitting contributions are absorbed into the 
bottom-quark PDF.   Alternative results with a central scale choice of $m_t/2$ in 
production, with the central scale $m_t$ retained in the decay part, show  
better convergence of the series although the NNLO predictions are almost unchanged.

In experimental analyses, the total inclusive cross sections are usually 
determined through extrapolation of the fiducial cross sections based on
acceptance estimates obtained from MC simulations.  We can use the numbers
shown in Tables~\ref{tab:total} and~\ref{tab:fiducial} to derive the
parton-level acceptance at various orders.  For top quark production, the
acceptances are 0.0283, 0.0214, and 0.0201 at LO, NLO, and NNLO respectively.
The NNLO corrections can change the acceptance by $6\%$ relative to the NLO value.  
This change also propagates
into the measurement of the total inclusive cross section through extrapolation.

To compare our results with those in Ref.~\cite{1404.7116}, we calculated the 
NNLO total inclusive cross sections at 8 TeV using the same choices of parameters.  
We found a difference of $\sim 1 \%$ on the NNLO cross sections.  
With a refined comparison through private communications, we traced the source of this 
discrepancy to NNLO contributions associated with $V_h$, with the $b$-quark initial state.
All other parts in the NNLO corrections and all parts of the NLO contributions agree between 
the two results within numerical uncertainties.  It has not been possible to further pin down the differences.  We leave this issue for possible future investigation.

\section{Differential Distributions}
\begin{figure}[!h]
  \begin{center}
  \includegraphics[width=0.45\textwidth]{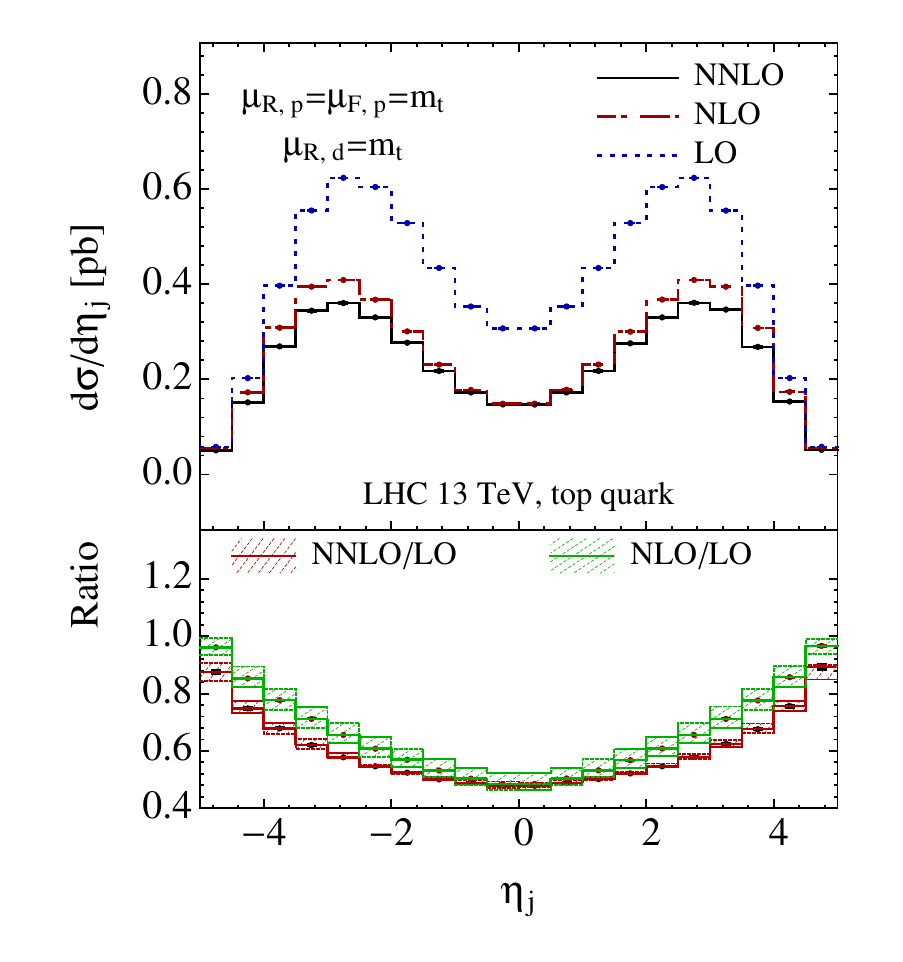}
  \end{center}
  \vspace{-2ex}
  \caption{\label{fig:pro13}
   Predicted pseudorapidity distribution of the non-$b$ jet in the final state 
   from top quark
   production with decay at 13 TeV with fiducial cuts applied.}
\end{figure}

\begin{figure}[!h]
  \begin{center}
  \includegraphics[width=0.45\textwidth]{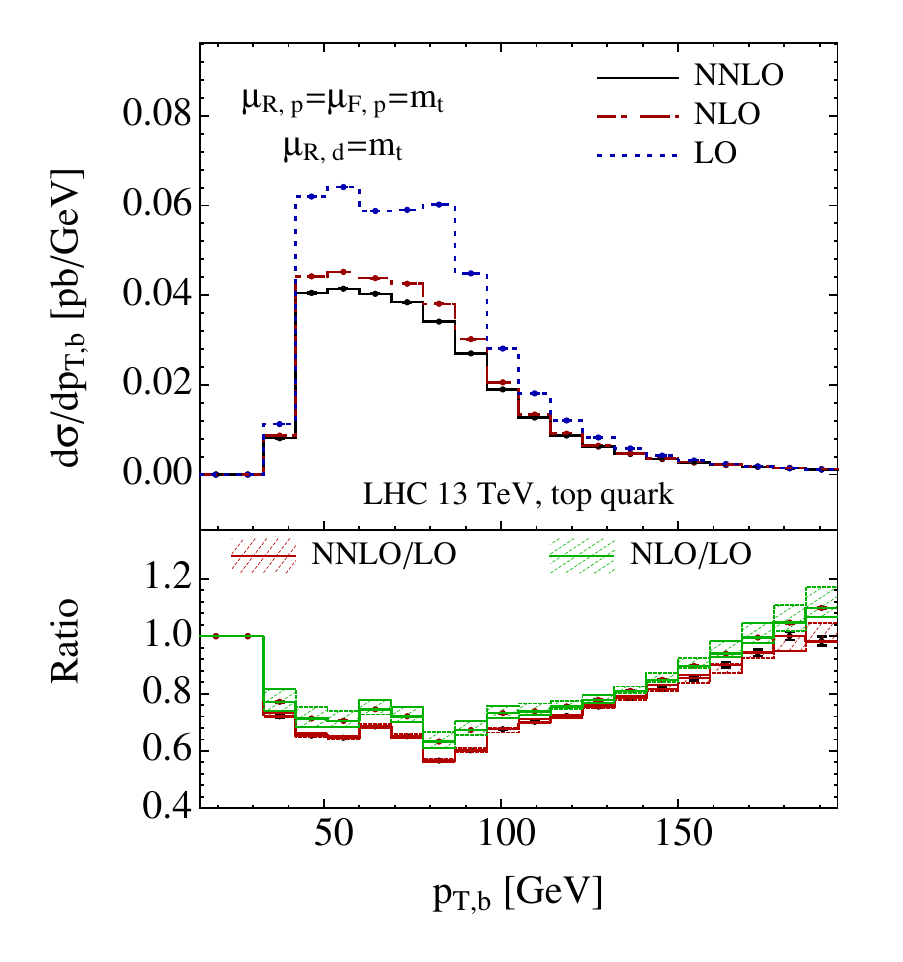}
  \end{center}
  \vspace{-2ex}
  \caption{\label{fig:dec06}
   Predicted transverse momentum distribution of the leading $b$-jet from top quark
   production with decay at 13 TeV with fiducial cuts applied.}
\end{figure}

We present differential distributions including NNLO corrections
for top quark production with decay. The effects for top anti-quark
distributions are similar. 
The full QCD corrections for 
the pseudorapidity distribution of the non-$b$ jet are shown in Fig.~\ref{fig:pro13} after all 
fiducial cuts are applied.  Events with two $b$-jets in the fiducial region are not included in the 
plot.  The corrections depend strongly on the pseudorapidity.  The 
NNLO corrections have a different shape from those at NLO and can be even larger than 
the NLO corrections in the regions of large pseudorapidity.  
The transverse momentum distribution of the leading $b$-jet is plotted in Fig.~\ref{fig:dec06},
again including the full QCD corrrections in production and decay.   
The corrections reach a maximum for $p_{T,b}$ of about 80 GeV.  Acceptance limitations explain 
the peculiar shape of the $p_T$ distribution.   We observe a reduction in the hard scale variations
in both Figs.~\ref{fig:pro13} and~\ref{fig:dec06}, calculated by varying the corresponding
scales in production and decay independently by a factor of two 
around $m_t$ and then adding the variations in quadrature.
In general we found large NLO corrections to the fiducial distributions, which makes
our NNLO calculation a necessity in order to assess the convergence and reliability of
pQCD series.

As mentioned in Sec.~\ref{sec:med}, we neglect cross-talk between incoming protons,  
and we discuss the justification for this approximation for the inclusive cross section.  Exchanges 
associated with cross-talk, although suppressed by a factor of $1/N_c^2$, might lead to different 
kinematical shape dependence for differential distributions, compared with the corrections considered 
in this manuscript.  It would be valuable to compute the cross-talk contributions in the future, once the 
relevant techniques are developed. We believe that the calculation presented in this manuscript
represents the best available results in the literature so far.

\begin{figure}[!h]
  \begin{center}
  \includegraphics[width=0.45\textwidth]{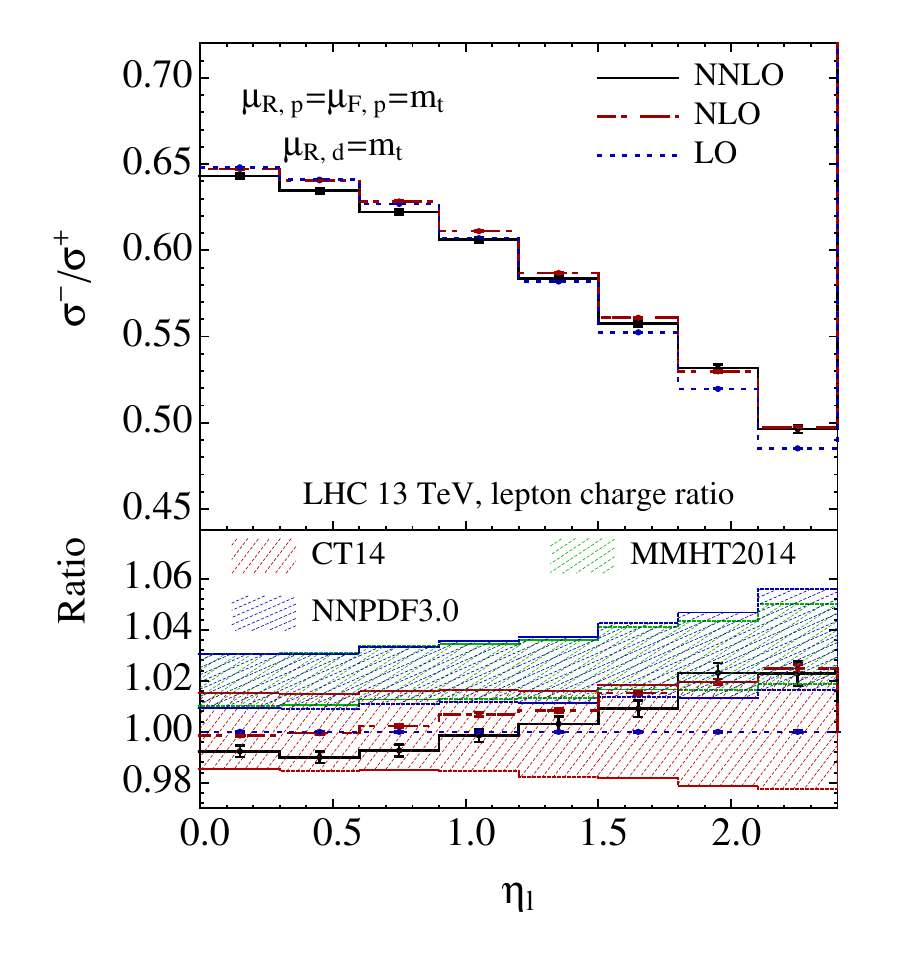}
  \end{center}
  \vspace{-2ex}
  \caption{\label{fig:ratio11}
  Ratios of the fiducial cross sections of top anti-quark to top quark production with decay
  at 13 TeV as a function of the pseudorapidity of the charged lepton.
  The lower panel shows ratios to the LO prediction as well as dependence on the choice of PDFs.}
\end{figure}

Charge asymmetry is one of the precision observables at the LHC,
e.g., as measured in $W$ boson production~\cite{Aaij:2012vn,Chatrchyan:2013mza,Aad:2011dm}.
It is insensitive to high-order
corrections and is less subject to experimental systematic uncertainties.  Moreover, since it is  
determined largely by the PDFs, it can provide stringent constraints in PDF 
determinations~\cite{Berger:1988tu,Dulat:2015mca}. 
The predicted ratio of the fiducial cross sections for 
top anti-quark and top quark production is presented in the upper panel of 
Fig.~\ref{fig:ratio11} as a function of the pseudorapidity of the charged lepton.
The ratio is less than one since there are more $u$-valence quarks than $d$-valence quarks 
in the proton, and it decreases with pseudorapidity because the $d/u$ ratio decreases at large
$x$~\cite{Dulat:2015mca}.  The uncertainty flags show the statistical uncertainty from the 
MC integration.  The ratios of the three curves are shown in the lower panel.  The spread of the LO, NLO, and 
NNLO predictions is about $1\%$ in the central region.  At large $|\eta_{l}|$, the NLO correction can reach 
about $2\%$, and the additional NNLO correction is well below one percent.  Also shown in the lower panel 
are the $68\%$ confidence-level uncertainty bands for three sets of NNLO PDFs: CT14~\cite{Dulat:2015mca}, 
MMHT2014~\cite{Harland-Lang:2014zoa} and NNPDF3.0~\cite{Ball:2014uwa}. 
For simplicity, we obtained these bands using the LO matrix elements and the NNLO PDFs, and we verified that 
quantitatively similar central values of the bands are obtained if we use NLO matrix elements.   
Since the PDF induced uncertainty is much 
larger than the theoretical uncertainty of its NNLO prediction, the charge ratio can be used 
reliably to further discriminate among and constrain the PDFs, provided that experimental
uncertainties can be controlled to the same level, as is also pointed out in~\cite{1404.7116,Alekhin:2013nda,Alekhin:2015cza}.
This charge ratio may also be sensitive to certain kinds of physics
beyond the SM~\cite{Gao:2011fx}.    

\section{Summary}
We present the first calculation of NNLO QCD
corrections to $t$-channel single top quark production with decay at the LHC
in the 5-flavor scheme in QCD, neglecting the cross-talk
between the hadronic systems of the two incoming protons.
Our calculation provides a fully differential 
simulation at NNLO for $t$-channel single top-quark
production with leptonic decay at the parton level.  The NNLO corrections
reduce the scale uncertainties of the theoretical predictions to a percent
level.  For the kinematic cuts used in the 8 TeV LHC experimenal analyses, 
the NNLO corrections to the fiducial cross sections can reach $-6$\%.  
Our results can be used to improve the determinations of the single top-quark 
production cross section and the top-quark electroweak coupling.\\

\begin{acknowledgments}
Work at ANL is supported in part by the U.S. Department of Energy
under Contract No. DE-AC02-06CH11357.
H.X.Z. was supported by the Office of Nuclear Physics of the U.S. DOE under Contract No. DE-SC0011090.
This research was supported in part by the National Science Foundation under Grant No. NSF PHY-1125915
and PHY-1417326.
We thank Ze Long Liu for cross-checking part of our results.
We thank K. Melnikov and F. Caola for providing numbers on inclusive  
cross sections for crosschecks. We also thank T. Gehrmann, A. Papanastasiou, A. Signer, 
and Z. Sullivan for useful conversations.
We thank Southern Methodist University for
the use of the High Performance Computing facility ManeFrame. \\
\end{acknowledgments}

\bibliographystyle{apsrev}
\bibliography{tchannel}

\begin{thebibliography}{60}
\bibitem[{foo()}]{foota}
\bibinfo{journal}{We define a parton-level $b$-jet as a clustered jet with
  non-zero bottom-quark number, e.g., a jet containing both a $b$ quark and
  anti $b$-quark is not considered as a $b$-jet.}
\end{thebibliography}
\end{document}